\documentstyle[preprint,aps,eqsecnum]{revtex}

\def\mycomm#1{\hfill\break\strut\kern-3em{\large \tt ====> #1}\hfill\break}
\def\mycommNL#1{\strut\kern-3em{\tt ====> #1}\hfill\break}
\def\ds{\displaystyle}
\def\deepstrut{\vrule height 1.5ex depth 3.5ex width 0pt}

\def\verydeepstrut{\vrule height 1.5ex depth 4.5ex width 0pt}
\def\tallstrut{\vrule height 4.5ex depth 1.5ex width 0pt}
\def\medstrut{\vrule height 2.5ex depth 1.3ex width 0pt}
\def\verytallstrut{\vrule height 6.5ex depth 1.5ex width 0pt}

\def\MUU{ {\cal V } }
\def\MUD{ {\cal P } }

\makeatletter
\def\hlinewd#1{\noalign{\ifnum0=`}\fi
\hrule \@height #1 \futurelet \reserved@a\@xhline}
\def\hwhiteline{\noalign
{\ifnum0=`}\fi\hrule 
\@height 0pt\vskip 1.0ex\futurelet \reserved@a\@xhline}
\makeatother
\def\gray{\special{ps: 0.40 setgray}}
\def\black{\special{ps: 0.0 setgray}}

\newcommand{\mydraft}{
\newcount\timecount
\newcount\hours \newcount\minutes  \newcount\temp \newcount\pmhours

\hours = \time
\divide\hours by 60
\temp = \hours
\multiply\temp by 60
\minutes = \time
\advance\minutes by -\temp
\def\hour{\the\hours}
\def\minute{\ifnum\minutes<10 0\the\minutes
    \else\the\minutes\fi}
\def\clock{
\ifnum\hours=0 12:\minute\ AM
\else\ifnum\hours<12 \hour:\minute\ AM
\else\ifnum\hours=12 12:\minute\ PM
    \else\ifnum\hours>12
	 \pmhours=\hours
	 \advance\pmhours by -12
	 \the\pmhours:\minute\ PM
	 \fi
    \fi
\fi
\fi
}
\def\fullclock{\hour:\minute}
\begin{centering}
\gray
\font\Hugett  =cmtt12 scaled\magstep4
\hbox{\Hugett Draft:\today,\clock}
\black
\end{centering}
\vskip -1.7cm
$\phantom{a}$
} 

\def\beq#1{\begin{equation} \label{#1}}
\def\eeq{\end{equation}}

\def\ket#1{\left\vert #1\right\rangle}

\newskip\humongous \humongous=0pt plus 1000pt minus 1000pt

\newif\ifdtup

\begin{document}
{\tighten
\preprint
{\vbox{\hbox{July 24, 2003}
\hbox{}
\hbox{TAUP 2735-03}
\hbox{hep-ph/0307243} }}

\title{The Constituent Quark Model Revisited - 
\\
Quark Masses, New Predictions for
Hadron Masses 
\\
and $KN$ Pentaquark}

\author{Marek Karliner\,$^a$\thanks{e-mail: \tt marek@proton.tau.ac.il}
\\
and
\\
Harry J. Lipkin\,$^{a,b}$\thanks{e-mail: \tt ftlipkin@clever.weizmann.ac.il}
}
\address{ \vbox{\vskip 0.truecm}
$^a\;$School of Physics and Astronomy \\
Raymond and Beverly Sackler Faculty of Exact Sciences \\
Tel Aviv University, Tel Aviv, Israel\\
\vbox{\vskip 0.0truecm}
$^b\;$Department of Particle Physics \\
Weizmann Institute of Science, Rehovot 76100, Israel \\ }
\maketitle

\begin{abstract}%
Prompted by the recent surprising results in 
QCD spectroscopy,
we extend to heavy flavors the hadron mass relations 
showing that the constituent quark mass differences and ratios have
the same values when  obtained from mesons and baryons.
We obtain several new successful relations involving heavy quarks and provide some
related predictions.
We discuss in detail the apparent sharp decrease in $m_s$ and $m_c$,
when a light partner quark in a meson is replaced by a heavy
one and construct a potential model which qualitatively reproduces 
this pattern through wave function effects. We apply these ideas to the recently
discovered $\Theta^+$ exotic $KN$ resonance 
and propose its interpretation as a novel
kind of a pentaquark with an unusual color structure, $J^P=1/2^+$, $I=0$ and
an antidecuplet of $SU(3)_f$.
A~rough mass estimate
of this pentaquark is close to experiment.
\end{abstract}%
} 

\section{Introduction }

Recently there have been several new surprising experimental
results in QCD spectroscopy:
enhancements near ${\bar p} p$ thresholds~\cite{BES,Belle}, an
exotic 5-quark $K N$ resonance \cite{Kyoto,Russia,Stepanyan:2003qr}, 
and two new extremely narrow mesons
containing $c$ and ${\bar s}$ quarks~\cite{Aubert:2003fg,CLEO,BELLE-Ds}.

These results triggered a vigorous theoretical activity 
and
put a renewed urge in the need to refine our quantitative 
understanding of how baryon and meson properties are obtained from quarks.

Early evidence that mesons  and
baryons are made of the same quarks was provided by the remarkable
successes of the constituent quark model, in which static properties
and low lying excitations of both mesons and baryons are described
as simple composites of asymptotically free quasiparticles 
with given effective masses.
For example,
the effective quark mass difference $m_s-m_u$
is found to have the same value $\pm 3\%$ and the mass ratio $m_s/m_u$
the same value $\pm 2.5\%$, when calculated  from baryon masses and from
meson masses\cite{SakhZel,ICHJLmass,HJLMASS}.

As long as QCD calculations 
have not yet succeeded to explain these striking
experimental facts from first principles, 
it is of interest to extend our phenomenological understanding 
of these phenomena.

\bigskip
\centerline{ \bf Relations between masses of baryons and mesons containing
light quarks}  

Andrei Sakharov, a pioneer in quark-hadron physics asked in 1966 ``Why are the
$\Lambda$ and $\Sigma$ masses different? They are made of the same quarks".
Sakharov and  Zeldovich\cite{SakhZel}  assumed a  quark model for hadrons with
a flavor dependent linear mass term and hyperfine interaction, 

\beq {sakzel}
M = \sum_i m_i + \sum_{i>j}  {{\vec{\sigma}_i\cdot\vec{\sigma}_j}\over{m_i\cdot
m_j}}\cdot v^{hyp}_{IE} 
\end{equation} 
\
where $m_i$ is the
effective mass of quark $i$, $\vec{\sigma}_i$ is a quark  spin operator and
$v^{hyp}_{ij}$ is a hyperfine interaction with different strengths but the same
flavor dependence.

Using (\ref{sakzel})
Sakharov and Zeldovich noted that 
both the mass difference $m_s-m_u$ between strange and nonstrange quarks and
the flavor dependence of their hyperfine splittings 
(later related\cite{DGG} to the mass ratio $m_s/m_u$) have the same values 
when calculated from baryon masses and meson masses\cite{SakhZel},
along with the comment that the masses
are of course effective masses\cite{Postcard}:
$\deepstrut$

\def\mykern{\kern-0.3em}
\begin{equation}
\kern-0.5em
\matrix{
\langle m_s{-}m_u \rangle_{Bar}
\mykern 
&=& 
\mykern 
M_{sud}{-}M_{uud}
\mykern 
&=& 
\mykern 
M_\Lambda{-}M_N
\mykern
&=&
\mykern
177\,{\rm MeV} \deepstrut
\cr 
\langle m_s{-}m_u \rangle_{Mes} 
\mykern
&=&
\mykern 
\ds
{
3  (M_{\MUU_{s\bar d}} {-} M_{\MUU_{u\bar d}})
{+}(M_{\MUD_{s\bar d}} {-} M_{\MUD_{u\bar d}})
\over 4}
\mykern 
&=&
\mykern 
\ds { 3(M_{K^{\scriptstyle *}}{-}M_\rho){+}M_K{-}M_\pi \over 4 }
\mykern 
&=&
\mykern 
179\,{\rm MeV}
\verydeepstrut
}
\label{SZeq}
\end{equation}
\vskip 10em
\begin{equation}
\left({{m_s}\over{m_u}}\right)_{Bar} =
{{M_\Delta - M_N}\over{M_{\Sigma^*} - M_\Sigma}} = 1.53 
\approx
\left({{m_s}\over{m_u}}\right)_{Mes} =
{{M_\rho - M_\pi}\over{M_{K^*}-M_K}}= 1.61
\verydeepstrut
\end{equation}
where the {\em ``Bar"} and {\em ``Mes"} subscripts denote values obtained
from baryons and mesons, and
$\MUU$ and $\MUD$ denote vector and pseudoscalar mesons, respectively.

Sakharov and Zeldovich isolated the contributions of the two terms in  (\ref
{sakzel}) by choosing mass differences that cancel contributions of either the
additive first term or the second hyperfine term. They canceled the  hyperfine
term in baryons by noting that the hyperfine interactions in the $N$ and
$\Lambda$ are determined entirely by the $u-d$ interactions and drop out of
the $\Lambda$ nucleon mass difference. In mesons they assumed the spatial wave
functions for the vector and pseudoscalar mesons to be the same and chose  the
linear combination which canceled the expectation value of the hyperfine term. 
Their relations do not require any relations between the 
space wave functions of the mesons, octet baryons and decuplet baryons.
We shall see later experimental hints of small 
differences between the octet and and decuplet space wave functions.

In this paper we generalize the Sakharov-Zeldovich approach to other flavors.    
Thus, given the masses of two vectors
\ $|\MUU_i\rangle = | q_i \bar x\rangle^{J=1}$
\ and
\ $|\MUU_j\rangle = | q_j \bar x\rangle^{J=1}$,
\ as well as the masses of the corresponding pseudoscalars,
\ $|\MUD_i\rangle = | q_i \bar x\rangle^{J=0}$
\ and \
$|\MUD_j\rangle = | q_j \bar x\rangle^{J=0}$,
\ we have, in analogy with the second equation in (\ref{SZeq}),
\begin{equation}
\langle m_{q_i} -m_{q_j} \rangle_{xMes}
=
{3(M_{\MUU_i} - M_{\MUU_j})
+(M_{\MUD_i} - M_{\MUD_j})
\over 4}
\end{equation}

This method fails for the $s \bar s$ system because of $\eta$-$\eta^\prime$ 
mixing and the absence of a pseudoscalar $\bar s s$ meson. 

For baryons we consider only the nucleon and the isoscalar baryons with one
heavy quark, $\Lambda$, $\Lambda_c$ and $\Lambda_b$,  where the hyperfine
interaction is determined entirely by the light quark $u$ and $d$ interactions
and drops out of all mass differences considered.

Thus given the masses of two baryons
\ $|B_i\rangle = | q_i ud \rangle$ \ and
\ $|B_j\rangle = | q_j ud \rangle$, \
we have
\begin{equation}
\langle m_{q_i} -m_{q_j} \rangle_{dBar}
= M_{q_i u d} - M_{q_j u d}
= 
M_{B_i} - M_{B_j}
\end{equation}
For any other two baryon states \ $|B_i\rangle = | q_i xy \rangle$ \ and \
$|B_j\rangle = | q_j xy \rangle$ \ with flavors $x$ and $y$ not related by
isospin, the spin of the $xy$ pair in any mass eigenstate is an unknown 
mixture of 0 and 1 and  the hyperfine interaction cannot be cancelled.

These mass relations are obtained assuming that the spatial wave functions for
the vector and pseudoscalar mesons are the same so that the values for the
hyperfine splittings are easily obtained from the mass values. However the 
Sakharov-Zeldovich relations do not require any relations between the 
space wave functions of the mesons, octet baryons and decuplet baryons.
As we shall see later, there are experimental hints of small 
differences between the octet and and decuplet space wave functions.

Further extension of this approach led to two more relations for $m_s-m_u$,
when calculated from baryon masses and meson masses\cite{ICHJLmass,HJLMASS}, 
and to three magnetic moment predictions with no free
parameters\cite{DGG,Protvino},
\begin{equation}
\matrix{
\langle m_s-m_u \rangle_{Mes} &=&
\ds {{3 M_\rho\tallstrut + M_\pi}\over 8}
\cdot
\left({{M_\rho - M_\pi}\over{M_{K^*}-M_K}} - 1 \right)
&=& 180\,{\rm MeV}
\verydeepstrut
\cr
\langle m_s-m_u \rangle_{Bar} &=&
\ds {{M_N+M_\Delta}\over 6}\cdot
\left({{M_{\Delta}-M_N}\over
{M_{\Sigma^{\scriptstyle *}}-M_\Sigma}} - 1 \right)
&=&192\,{\rm MeV}
\verydeepstrut
}
\end{equation}
\begin{equation}
\verytallstrut
\mu_\Lambda=
-{\mu_p\over 3}\cdot {{m_u}\over{m_s}} =
-{\mu_p\over 3}\cdot {{M_{\Sigma^*} - M_\Sigma} \over{M_\Delta - M_N}}
=-0.61 \,{\rm n.m.}
\qquad(\hbox{EXP}  =-0.61 \,{\rm n.m.})
\end{equation}

\begin{equation}
{\mu_p \over \mu_n} =
-{3 \over 2}
\qquad(\hbox{EXP}  = -1.46 )
\end{equation}

\beq{isomag}
\mu_p+\mu_n= 
2M_{\scriptstyle p}\cdot {{Q_I}\over{M_I}} 
={2M_N\over M_N+M_\Delta}=0.865 \,{\rm n.m.}
\qquad(\hbox{EXP}  =
0.88 \,{\rm n.m.})
\verydeepstrut
\end{equation}
where   
$ Q_I= \ds {1\over 2}\cdot \left( {2\over3} - {1\over 3} \right) =
 {1\over 6} $ \ and \
$M_I= \ds {1\over 6}\cdot \left( M_N + M_\Delta \right)$
denote the charge and mass, respectively, of an effective 
``isoscalar nonstrange quark". Note the implicit assumption
that in $M_I$ the contribution
of the hyperfine interaction is cancelled between the nucleon
and the $\Delta$.
The same approach has been applied to $m_b-m_c$ 
\cite{PBIGSKY} with excellent results (see Table I below).
Our quantitative understanding of QCD is not yet sufficient to understand 
why the simple constituent quark model is so 
remarkably  successful.

\section {New relations between meson and baryon masses from hadrons
containing heavy quarks.}

\subsection {Application of the Sakharov-Zeldovich model to  hadrons
containing heavy quarks.}

The availability of data giving masses of hadrons containing heavy quarks
enables us to push this model further to see where it works and where it breaks
down. 

The following Table lists the effective quark mass differences obtained 
from baryons and mesons.
We consider here the mass differences between two hadrons containing 
a quark of flavor $i$ and a quark of flavor $j$, bound to a common 
``spectator"
(anti)quark or diquark of
flavor $x$. We use the notation $\langle m_i-m_j \rangle_{x}$. 
For light spectator quarks $x=d$ denotes both $d$ and $u$ quarks.

\vskip1.5cm
\vbox{
{\def\baselinestretch{1.1}\small\normalsize
\centerline{{\bf TABLE I - Quark mass differences from baryons and mesons}}
\hfill\break
$$
\begin{array}{|c||c|c||c|c|c|c||c|c|}
\hline
&\multicolumn{2}{c||}{}&\multicolumn{4}{c||}{\hbox{mesons}}&&
\\
\cline{4-7}
&\multicolumn{2}{c||}
{\raisebox{2.5ex}[0pt]{\hbox{baryons}}}
&\multicolumn{2}{c|}{J=1}&\multicolumn{2}{c||}{J=0}
&\Delta m_{Bar} & \Delta m_{Mes}
\\
\cline{2-7}
\raisebox{5.0ex}[0pt]{\hbox{observable}}
& B_i  & B_j & \MUU_i & \MUU_j & \MUD_i & \MUD_j 
&\hbox{MeV} & \hbox{MeV}
\\
\hline
\hwhiteline
\hline
& sud  & uud & s\bar d & u\bar d& s \bar d & u \bar d&&\\
\cline{2-7}
\raisebox{2.5ex}[0pt]
{\hbox{$\langle m_s-m_u \rangle_d$}}&\Lambda & N & K^*& \rho & K & \pi 
& \raisebox{2.5ex}[0pt]{177} 
& \raisebox{2.5ex}[0pt]{179} 
\\
\hline
\hline
&   &  & c\bar s & c\bar u& c \bar s & c \bar u&&\\
\cline{2-7}
\raisebox{2.5ex}[0pt]
{\hbox{$\langle m_s-m_u \rangle_c$}}& &  & D^*_s& D^*_s & D_s & D_s
& \raisebox{2.5ex}[0pt]{} 
& \raisebox{2.5ex}[0pt]{\,\,\,103} 
\\
\hline
\hline
&   &  & b\bar s & b\bar u& b \bar s & b \bar u&&\\
\cline{2-7}
\raisebox{2.5ex}[0pt]
{\hbox{$\langle m_s-m_u \rangle_b$}}& &  & B^*_s& B^*_s & B_s & B_s
& \raisebox{2.5ex}[0pt]{} 
& \raisebox{2.5ex}[0pt]{\,\,\,\,91} 
\\
\hline
\hwhiteline
\hline
& cud  & uud & c\bar d & u\bar d& c \bar d & u \bar d&&\\
\cline{2-7}
\raisebox{2.5ex}[0pt]
{\hbox{$\langle m_c-m_u \rangle_d$}}&\Lambda_c & N & D^*& \rho & D & \pi 
& \raisebox{2.5ex}[0pt]{1346} 
& \raisebox{2.5ex}[0pt]{1360} 
\\
\hline
\hline
&   &  & c\bar c & u\bar c& c \bar c & u \bar c&&\\
\cline{2-7}
\raisebox{2.5ex}[0pt]
{\hbox{$\langle m_c-m_u \rangle_c$}}& &  & \psi& D^* & \eta_c & D
& \raisebox{2.5ex}[0pt]{} 
& \raisebox{2.5ex}[0pt]{1095} 
\\
\hline
\hwhiteline
\hline
& cud  & sud & c\bar d & s\bar d& c \bar d & s \bar d&&\\
\cline{2-7}
\raisebox{2.5ex}[0pt]
{\hbox{$\langle m_c-m_s \rangle_d$}}&\Lambda_c & \Lambda & D^*& K^* & D & K
& \raisebox{2.5ex}[0pt]{1169} 
& \raisebox{2.5ex}[0pt]{1180} 
\\
\hline
\hline
&   &  & c\bar c & s\bar c& c \bar c & s \bar c&&\\
\cline{2-7}
\raisebox{2.5ex}[0pt]
{\hbox{$\langle m_c-m_s \rangle_c$}}&& & \psi& D^*_s & \eta_c & D_s
& \raisebox{2.5ex}[0pt]{} 
& \raisebox{2.5ex}[0pt]{991} 
\\
\hline
\hwhiteline
\hline
& bud  & uud & b\bar d & u\bar d& b \bar d & u \bar d&&\\
\cline{2-7}
\raisebox{2.5ex}[0pt]
{\hbox{$\langle m_b-m_u \rangle_d$}}& \Lambda_b & N & B^*& \rho & B & \pi
& \raisebox{2.5ex}[0pt]{4685} 
& \raisebox{2.5ex}[0pt]{4700} 
\\
\hline
\hline
&   &  & b\bar s & u\bar s& b \bar s & u \bar s&&\\
\cline{2-7}
\raisebox{2.5ex}[0pt]
{\hbox{$\langle m_b-m_u \rangle_s$}}& &  & B_s^*& K^* & B_s & K
& \raisebox{2.5ex}[0pt]{} 
& \raisebox{2.5ex}[0pt]{4613} 
\\
\hline
\hwhiteline
\hline
& bud  & sud & b\bar d & s\bar d& b \bar d & s \bar d&&\\
\cline{2-7}
\raisebox{2.5ex}[0pt]
{\hbox{$\langle m_b-m_s \rangle_d$}}&\Lambda_b & \Lambda & B^*& K^* & B & K
& \raisebox{2.5ex}[0pt]{4508} 
& \raisebox{2.5ex}[0pt]{4521} 
\\
\hline
\hwhiteline
\hline
& bud  & sud & b\bar d & c\bar d& b \bar d & c \bar d&&\\
\cline{2-7}
\raisebox{2.5ex}[0pt]
{\hbox{$\langle m_b-m_c \rangle_d$}}&\Lambda_b & \Lambda_c & B^*& D^* & B & D
& \raisebox{2.5ex}[0pt]{3339} 
& \raisebox{2.5ex}[0pt]{3341} 
\\
\hline
\hline
&   &  & b\bar s & c\bar s& b \bar s & c \bar s&&\\
\cline{2-7}
\raisebox{2.5ex}[0pt]
{\hbox{$\langle m_b-m_c \rangle_s$}}& &  & B^*_s& D^*_s & B_s & D_s
& \raisebox{2.5ex}[0pt]{} 
& \raisebox{2.5ex}[0pt]{3328} 
\\
\hline
\end{array}
$$ }}
The mass difference between two quarks of different flavors denoted by $i$ and
$j$ are seen to have the same value to a good approximation when they are bound
to a nonstrange antiquark to make a meson and  bound to a
nonstrange diquark to make a baryon,
\begin{equation}
\langle m_i-m_j \rangle_{dBar} \approx 
\langle m_i-m_j \rangle_{dMes}.
\end{equation}
We now calculate 
the mass ratio \ $m_c/m_s$ \ from the hyperfine splittings 
in mesons and baryons in the same way that  
the mass ratio $m_s/m_u$ has been calculated\cite{SakhZel,DGG}. 
We find the same value from mesons and baryons $\pm2\%$\,:
\begin{equation}
\left({{m_c}\over{m_s}}\right)_{Bar} =
{{M_{\Sigma^*} - M_\Sigma}\over{M_{\Sigma_c^*} - M_{\Sigma_c}}} = 2.84 = 
\left({{m_c}\over{m_s}}\right)_{Mes} =
{{M_{K^*}-M_K}\over{M_{D^*}-M_D}}= 2.81
\end{equation}
\begin{equation}
 \left({{m_c}\over{m_u}}\right)_{Bar} =
{{M_\Delta - M_p}\over{M_{\Sigma_c^*} - M_{\Sigma_c}}} = 4.36 = 
\left({{m_c}\over{m_u}}\right)_{Mes} =
{{M_\rho-M_\pi}\over{M_{D^*}-M_D}}= 4.46
\end{equation}
The presence of a fourth flavor gives us the possibility of obtaining a new 
type of mass relation between mesons and baryons. The $\Sigma - \Lambda$ mass
difference is believed to be due to the difference between the $u-d$ and $u-s$
hyperfine interactions. Similarly, the $\Sigma_c - \Lambda_c$ mass
difference is believed to be due to the difference between the $u-d$ and $u-c$
hyperfine interactions. We therefore obtain the relation 
\beq{SigLam4}
\vrule height 9.5ex depth 1.5ex width 0pt
\kern-1ex
\left({
\ds {1\over m_u^2} - {1\over m_u m_c}  
\over
\ds  {1\over m_u^2} - {1\over m_u m_s}}\right)_{\strut \kern-1ex Bar} 
\kern-3.0ex
={{M_{\Sigma_c} - M_{\Lambda_c}}\over{M_{\Sigma} - M_\Lambda}}=2.16
\approx
\left({
\ds {1\over m_u^2} - {1\over m_u m_c}  
\over
\ds  {1\over m_u^2} - {1\over m_u m_s}}\right)_{\strut \kern-1exMes} 
\kern-3.0ex
=
{{(M_\rho {-} M_\pi){-}(M_{D^*}{-}M_D)}
\over
{(M_\rho {-} M_\pi){-}(M_{K^*}{-}M_K)}}
=2.10
\end{equation}
The meson and baryon relations agree to $\pm 3\%$.

\noindent
We also obtain a similar relation for $\Lambda_b$ and the yet unmeasured 
$\Sigma_b$:
\begin{equation}
{{M_{\Sigma_b} - M_{\Lambda_b}}\over{M_{\Sigma} - M_\Lambda}} = ? = 
{{(M_\rho - M_\pi)-(M_{B^*}-M_B)}\over{(M_\rho - M_\pi)-(M_{K^*}-M_K)}}= 2.51
\end{equation}
This then predicts that $M_{\Sigma_b} = 5818 \,{\rm MeV}$  and 
 $M_{\Sigma_b} - M_{\Lambda_b} = 194 \,{\rm MeV}$.

These results have been obtained without any explicit model for the hyperfine 
interaction beyond the flavor dependence of the $\Sigma-\Lambda$ 
mass difference.

\pagebreak
\subsection {Summary of successful mass relations from hadrons 
containing no more than one strange or heavy quark.}  

The effective quark mass appears in two terms in the mass
formula (\ref{sakzel}) -- as an additive term and in the denominator of the
hyperfine interaction. In all the relations for masses and magnetic moments
obtained in the light ($uds$) 
flavor sector, agreement with experiment has been obtained
by assuming that the values of the effective quark masses in these two terms
has been the same and that the values are the same for mesons and baryons.
Both the mass difference and the mass ratio between two quarks of different 
flavors denoted by $i$ and $j$ are seen to have the same values to a good
approximation when they are bound to a nonstrange antiquark to make a meson
and  bound to a nonstrange diquark to make a baryon. However differences arise
when we examine mass differences between quarks bound in mesons to a strange or 
heavier antiquark. It is already remarkable that this simple description is so
successful, since these are effective masses which include the kinetic and
potential energies of the hadron. It is not obvious how these energies are
divided among the effective masses of the individual quarks having different
flavors in a two-body or three-body system. 

We test this picture further by attempting to fit both the mass differences and
mass ratios with a single set of quark masses. 
We choose
\beq{qmass}
m_u =  360  
\hbox{\ MeV};\qquad
m_s=  540 
\hbox{\ MeV};\qquad
m_c= 1710 
\hbox{\ MeV};\qquad
m_b=  5050
\hbox{\ MeV}\,.
\end{equation}
These have been chosen to give an eyeball fit to the baryon mass differences
and to fit the isoscalar nonstrange baryon magnetic moment
\beq{qmag} 
\mu_p+\mu_n= 0.88 \,{\rm n.m.} ={M_N\over 3m_u}
=0.87 \,{\rm n.m.}
\end{equation}
The results are shown in Table II below. 

\vskip1cm
\vbox{
{\centerline{\bf{Theoretical and Experimental Hadron Mass Differences and Ratios}}}

\vskip 1.truecm

\centerline{\bf{TABLE II-A - Hadron Mass Differences}}
$$ \vcenter{
\halign{${#}$\quad
        &${#}$\quad
        &${#}$\quad        	
        &${#}$\cr
 {\rm \ \ \ Mass~Difference} &{\rm Theoretical}&{\rm Experimental}&{\rm Experimental} \cr
    & {\rm From~eq.~(\ref{qmass})}   & {\rm From~Mesons}~(X{=}d) & {\rm From  ~ Baryons}~(X{=}ud)\cr
m_s{-}m_u = M(sX) {-} M(uX)\ \  &  \,\,180  & \,\, 179   &   \,\,  177     \cr
m_c{-}m_u = M(cX) {-} M(uX) & 1350  & 1360  &   1346    \cr
m_b{-}m_u = M(bX) {-} M(uX) & 4690 &  4701   &    4685    \cr
m_c{-}m_s = M(cX) {-} M(sX) & 1170 &  1180   &  1169   \cr
m_b{-}m_s = M(bX) {-} M(sX) & 4510 &  4521  &     4508   \cr
m_b{-}m_c = M(bX) {-} M(cX) & 3340  & 3341  &   3339     \cr
}}   $$ 
}
\par

\vskip 2.truecm

\centerline{\bf{TABLE II-B - Quark Mass Ratios}}
\nopagebreak
$$ \vcenter{
\halign{${#}$\quad
        &${#}$\quad
        &${#}$\quad        	
        &${#}$\cr
 {\rm Mass~Ratio} & {\rm Theoretical} & {\rm Experimental}& {\rm Experimental} \cr
   & {\rm From~eq.~(\ref{qmass})}  & {\rm From~Mesons}~(X=d)&{\rm From~Baryons}~ (X=ud)\cr
\ m_s/m_u &  1.5   &   1.61 &   1.53      \cr
\ m_c/m_u &  4.75   &  4.46  &  4.36 \cr
\ m_b/m_u &  14.0  & 13.7  &   ?    \cr
\ m_c/m_s &  3.17  & 2.82  &  2.82  \cr
\ m_b/m_s &  9.35 &  8.65  &  ?\cr
\ m_b/m_c &  2.95  & 3.07  &  ?  \cr
}}   $$ 
\hfill\break
\par 
The results shown in Table II for the mass differences simply express the
close agreement between meson and baryon entries in Table I.
The results for the mass
ratios test the equality of the effective masses obtained from 
the additive term and from the hyperfine term
in the mass formula. 
The overall agreement between the experimental and theoretical 
quark mass ratios in Table II-B is still quite good,
better than 10\%.
However, the the experimental meson and baryon
quark mass ratios obtained from the hyperfine term
are much closer to each other than to 
theoretical predictions using quark masses
extracted from the additive term. 
At present we do not have a good understanding of the origin and sign of
these small deviations.

\section{$SU(6)$ breaking and effects of confining potential}
We now attempt to go beyond the simplest good relations,
yet unexplained as yet by QCD, and look into further relations, as well as
investigate the behavior of hadrons in which more than one quark is strange or
heavier.

\subsection {Improving the Sakharov-Zeldovich model by breaking $SU(6)$}
When the simple quark model is used to relate quark spin couplings in baryons,
the $\Sigma-\Lambda$ mass difference can be directly related to masses
of other baryons.
\beq{SigLam}
M_\Sigma -   M_\Lambda= 77  \,{\rm MeV} 
\quad\approx\quad
 {{2}\over{3}} \cdot 
\left[ (M_\Delta - M_p) - (M_{\Sigma^*} - M_\Sigma)\right] =  67 \,{\rm MeV}
\end{equation}
The difference between the $u$-$d$ and $u$-$s$
hyperfine interactions is proportional to \
\hbox{$(M_\Delta - M_p) - (M_{\Sigma^*} - M_\Sigma)$} \ \ and the factor 
$(2/3)$ is obtained from the quark spin couplings. 

The prediction
in eq.~(\ref{SigLam}), which assumes specific quark spin couplings, is not in
as good agreement with experiment as the other successful predictions
including  (\ref{SigLam4}) for $\Sigma - \Lambda$ splittings.  One possible
explanation is the breakdown of the $SU(6)$ symmetry  which assumes the same
space part in the baryon octet and decuplet wave functions, typically both
$S$-wave. This assumption has not been needed elsewhere in this paper, except
for the magnetic moment relation (\ref{isomag}) which is discussed below.  It
is used here to relate the  explicit values of $(M_\Delta - M_p)$ and
$(M_{\Sigma^*} - M_\Sigma)$ to the  values of the individual hyperfine
interactions in the baryon octet. The strengths of both hyperfine interactions
are then given by the expectation value of the relevant operators in the 
same space part of the wave function. 

We investigate the possible breaking of $SU(6)$ by 
introducing a symmetry breaking parameter $\xi$ relating the 
octet ({\bf 8}) and decuplet ({\bf 10})
expectation values of the hyperfine interactions in eq. (\ref{sakzel})
  
\beq {sakzel1}
\sum_{i>j}  
\left\langle{{\vec{\sigma}_i\cdot\vec{\sigma}_j}\over{m_i\cdot
m_j}}\cdot v^{hyp}_{ij}\right\rangle_{\hbox{\bf 10}}
\,=\,
 - (1 - 2\xi) \,\sum_{i>j}  
\left\langle{{\vec{\sigma}_i\cdot\vec{\sigma}_j}\over{m_i\cdot
m_j}}\cdot v^{hyp}_{ij}\right\rangle_{\hbox{\bf 8}}
\end{equation} 
where in the parameter $\xi$  measures the deviation 
of the baryon wave functions from $SU(6)$ symmetry,
so that $\xi=0$ corresponds to the
$SU(6)$ symmetric limit.
\beq {sakzel2}
M_\Delta -M_N = - 2(1 - \xi) 
\,\sum_{i>j} \left \langle{{\vec{\sigma}_i\cdot\vec{\sigma}_j}\over{m_i\cdot
m_j}}\cdot v^{hyp}_{ij}\right\rangle_{\hbox{\bf 8}}
\end{equation} 
This provides the correction factor for the $\Sigma - \Lambda$ relation 
(\ref{SigLam})
 
We now 
fix $\xi$ to fit the experiment and then look for other places where
$\xi$ can affect predictions, in particular
the isoscalar nucleon magnetic moment. Given
\begin{equation}
M_\Sigma -   M_\Lambda= 77  \,{\rm MeV} = {{2}\over{3\cdot (1 - \xi)}}  
\left[ (M_\Delta - M_p) - (M_{\Sigma^*} - M_\Sigma)\right] = 
{{ 67}\over{(1-\xi) }}\,{\rm MeV}
\end{equation}
This will agree with experiment if 
\begin{equation}
1 - \xi = 67/77 \qquad\hbox{or}\qquad \xi = 0.13\,.
\label{xivalue}
\end{equation}
To examine further case where this correction arises we note that
\beq {sakzel7}
M_\Delta +M_N  = 6 M_I + 2 \xi\cdot \sum_{i>j} 
\left \langle{{\vec{\sigma}_i\cdot\vec{\sigma}_j}\over{m_i\cdot
m_j}}\cdot v^{hyp}_{ij}\right\rangle_{\hbox{\bf 8}} = 
6 M_I - {{ \xi}\over{1 - \xi}}\cdot(M_\Delta -M_N)
\end{equation} 
\beq   {sakzel8}
 M_I  = \left({{ M_\Delta +M_N}\over{6}} \right)\cdot 
 \left( 1 + {{ \xi}\over{1 - \xi}}\cdot
 {{M_\Delta -M_N}\over{M_\Delta +M_N}}\right)
\end{equation} 
where $M_I$ denotes the mass of the effective ``isoscalar nonstrange quark".
used in eq. (\ref{isomag}).
 
Note that in $M_I$ the contribution
of the hyperfine interaction is cancelled between the nucleon
and the $\Delta$ when $SU(6)$ wave functions are used.  
If we choose 
$\xi = 0.13$, 
so that the $SU(6)$ breaking fits the  $\Sigma-\Lambda$ difference, then 
$${{\xi}\over{1 - \xi}}\cdot {{M_\Delta -M_N}\over{M_\Delta +M_N}}= 0.02$$
and
\beq{isomag_corr}
2M_{\scriptstyle p}\cdot {{Q_I}\over{M_I}} 
={2M_{\scriptstyle p}\over M_N+M_\Delta}\cdot \left( 1 + 
{{ \xi}\over{1 - \xi}}\cdot {{M_\Delta -M_N}\over{M_\Delta +M_N}}\right)^{-1}
=0.88 \,{\rm n.m.}\qquad ({\rm EXP} = 0.88 \,{\rm n.m.})
\end{equation}
to be compared with 0.865\,n.m. in the symmetry limit, cf. (\ref{isomag}).
It is interesting that even though $\xi=0.13$ the correction in 
(\ref{isomag_corr}) is small $\sim 1.7\%$.
That it is just in the right direction to even improve
the prediction may well be fortuitous.
 
We now examine the possible origin of the $SU(6)$ breaking in baryon wave
functions described by the phenomenological parameter $\xi$.  

The space parts of the  $N$ and the $\Delta$ wave functions  cannot be strictly the
same.  Their mass difference of about 300 MeV is of the order of the
constituent quark mass. So  some relativistic effects are expected.

Noting that even the nonrelativistic deuteron wave function contains a $\sim
5$\% $D$-wave admixture \cite{Machleidt:hj}, we look for a similar  $^3D_1$ -
$^3S_1$ quark pair mixing in the  decuplet where a $^3D_1$ pair can couple with
the third quark to a total $J=3/2$. It is forbidden by $SU(3)$ in the baryon
octet,  where the $L=2$ of the d-wave must combine with a total spin 3/2 of the
three quarks to make $J=1/2$, but three quarks coupled to $S = 3/2$ must be in
a decuplet of  $SU(3)$. ($P$-wave does not contribute, since it changes parity)

Another way to see it is to note that 3 quarks coupled to spin 3/2 can combine
with either  $L=0$ or $L=2$ to give a baryon with total $J=3/2$,  but 3 quarks
coupled to spin 1/2 cannot couple with $L=2$ to  give total $J=1/2$, so there
can be no $D$-wave admixture in the baryon octet. 
The quadrupole admixtures are currently being 
investigated experimentally\cite{jlab}.

$SU(3)$ breaking 
can mix the octet $\Sigma$ and $\Xi$ with their decuplet partners
$\Sigma^*$ and $\Xi^*$,
to produce a quark spin 3/2 and 
introduce quadrupole  admixture. 
But the $\Lambda$ and the nucleon cannot mix with any of the 
decuplet baryons, because of isospin conservation.

This may help to explain the robustness of the predictions based on the
assumption that the nonstrange  pair in the $\Lambda$ is coupled to spin zero,
the $\Lambda$ total angular momentum  and magnetic moment is due to the strange
quark and the hyperfine interactions are equal in the nucleon and $\Lambda$.

\subsection {Effects of the confining potential type on 
$\langle\MakeLowercase{m_s-m_u}\rangle$ }  
The mass difference between strange and nonstrange quarks
$m_s-m_u$ seems to have very different values when calculated naively from 
masses of mesons containing light ``spectator" quarks vs. those containing
heavy $b$ and $c$ ``spectator" quarks. In contrast to the value of
$179\,{\rm MeV}$ obtained for $\langle m_s-m_u\rangle_{dMes} $, we obtain
\begin{equation}
\verytallstrut
\langle m_s-m_u \rangle_{cMes} = {{3(M_{D_s^{\scriptstyle *}}
-M_{D^{\scriptstyle *}})
+M_{D_s}-M_D}\over 4} =103\,{\rm MeV} 
\end{equation}
\begin{equation}
\langle m_s-m_u \rangle_{bMes}= {{3(M_{B_s^{\scriptstyle *}}
-M_{B^{\scriptstyle *}})
+M_{B_s}-M_B}\over 4} = \,\,\,\,\,91\,{\rm MeV}
\end{equation}

To investigate this difference in more detail we first note that the effective
mass also includes the potential and kinetic energies of the quarks. These
depend upon the hadron wave function and can change from one hadron to
another.  Thus the difference between the effective masses of quarks of flavor
$i$ and $j$ can depend upon their environment; i.e.
\beq {secdiff}
\langle m_i-m_j \rangle_{xMes} - \langle m_i-m_j \rangle_{yMes} \not= 0. 
\end{equation}
where $x$ and $y$ are two different ``spectator" quarks, such that 
$m_x \ne m_y$.

To estimate this effect we are guided by the success of the simple
potential models used to describe the  charmonium and bottomonium spectra. We
rewrite the Sakharov-Zeldovich mass formula
(\ref{sakzel})
 for the case of a meson containing
a quark of flavor $i$ and an antiquark of flavor $j$, to include 
a nonrelativistic
potential model for the relative  quark-antiquark motion of the $q
\bar q$ pair described by an effective Hamiltonian $H_{ij}(m_r)$,
\beq {sakzelnu}
M(ij) =  m_i + m_j + {{\vec{\sigma}_i\cdot\vec{\sigma}_j}\over{m_i\cdot
m_j}}\cdot v^{hyp}_{ij}  =  m^c_i + m^c_j + H_{ij}(m_r) + 
{{\vec{\sigma}_i\cdot\vec{\sigma}_j}\over{m_i\cdot
m_j}}\cdot v^{hyp}_{ij}
\end{equation} 
where $m_r$ is the reduced mass of the quark-antiquark system obtained from
 the ``constituent quark masses" $m^c_i$ and $m^c_j$,
which are loosely defined
as what remains of the effective quark masses after the kinetic and potential
energies of the relative motion are removed. We assume that, unlike the
``total" masses $m_i$ and $m_j$, the
constituent masses do not depend
upon their environment and  therefore drop out of the second order differences
(\ref{secdiff}) that are of interest:
\beq {secdiff2}
\begin{array}{ccccc}
\langle m_i-m_j \rangle_{xMes} - \langle m_i-m_j \rangle_{yMes} &=&
 [M(ix) - M(jx)] &-& [M(iy) - M(jy)] 
\\
&=&
[E(ix) - E(jx)] &-& [E(iy) - E(jy)]
\end{array}
\end{equation}
where $E(ix)$ denotes  $\langle H_{ix}(m_r) \rangle$ and we average over the
hyperfine splitting, so the potential energy does not depend on the mass.
The effective Hamiltonian
$H_{ij}(m_r)$ depends upon the flavors $i$ and $j$ only via the reduced mass
$m_r$. To determine this dependence, we use  the
Feynman-Hellmann theorem  to obtain,
\begin{equation}
{{d}\over{dm_r}}\cdot \langle H_{ij}(m_r) \rangle
= \left\langle {{dH_{ij}(m_r)}\over{dm_r}} \right\rangle = 
- {{\langle  T \rangle_m}\over{m_r}}
\end{equation}
where 
$T=\ds{p^2\over2 m_r}$ denotes 
kinetic energy of the quark-antiquark 
system\footnote{\tighten The nonrelativistic expression for constituent quarks 
is a good approximation
when discussing the ground state, but relativistic corrections become
important for excited states.}
and $\langle  T \rangle_m$ 
denotes its expectation value with a wave function for a reduced mass $m_r=m$.

Integrating the equation gives for any two quark flavors $i$ and $j$ bound to a
common antiquark or diquark denoted by $x$,
\beq{FH1}
 E(ix) - E(jx)  
 = {-}\int_{m(jx)}^{m(ix)} dm_r 
\left\langle {{dH(m_r)}\over{dm_r}} \right\rangle 
 =  \int_{m(jx)}^{m(ix)} dm_r 
 {{\langle  T \rangle_m}\over{m_r}}
\end{equation}
This expression (\ref{FH1}) reduces to  a particularly simple form for the 
Quigg-Rosner 
logarithmic potential\cite{quiggros} which fits the charmonium and bottomonium spectra and is
particularly easy to use in calculations. 
\begin{equation}
V_{QR} = V_o \cdot \log \left({r\over r_o}\right); 
\qquad\langle T \rangle_{QR} = {1\over2} \,V_o
\end{equation}
where the parameter $V_o$ was determined by fitting the charmonium spectrum.
We then have
\begin{equation}
 E(ix) {-} E(jx) {=}
\int_{m(ix)}^{m(jx)} dm_r {{V_o}\over{2m_r}}
= {V_0\over2} \cdot \log \left [{{m_r(jx)}\over{m_r(ix)}} \right ]  
   = {V_0\over2} \cdot 
\log \left [{{m_j \cdot (m_i + m_x)}\over{m_i \cdot (m_j + m_x)}} \right ] 
\label{EixEjx}
\end{equation}
The values of $ E(ix) {-} E(jx)$ obtained from eq.~(\ref{EixEjx})
are listed in Table III below.
\hfill\break
\hfill\break
\vbox{
\begin{center}
{\bf TABLE III \\
Effective quark mass differences, depending on the
spectator}
\end{center}
{\def\baselinestretch{1.2}\small\normalsize
$$ \begin{array}{|c|c|c|c|}
\hline
\ \ \ q_i \ \ \ & \ \ \ q_j \ \ \ & \ \ \ q_x \ \ \ & \ E(ix) {-} E(jx) \    \\ 
    &     &     & \hbox{(MeV)}        \\ \hline
d   &  s  &  u  &  \,\,66                 \\ \hline
d   &  s  &  c  &  117                \\ \hline
d   &  s  &  b  &  136                \\ \hline
s   &  c  &  c  &  267                \\ \hline
u   &  c  &  c  &  384                \\ \hline
u   &  c  &  d  &  183                \\ \hline
s   &  c  &  d  &  116                \\ \hline
c   &  b  &  d  &   \,\,44                \\ \hline
c   &  b  &  s  &   \,\,63                \\ \hline
s   &  b  &  c  &  413                \\ \hline
s   &  c  &  b  &  350                \\ \hline
d   &  c  &  b  &  486                \\ \hline
s   &  b  &  d  &  161                \\ \hline
\end{array}$$
}
}
\hfill\break
\noindent  
The quark mass values (\ref{qmass}) were used and 
$\ds {V_0\over 2} = 364$
was taken from the Quigg-Rosner fit to charmonium \cite{quiggros}. 
Clearly, not all the numbers in Table III are independent, 
for example 
$384 = 117 +267$, \
$350 = 413 - 63$, \
$486 = 136 +350$, etc. 
We list them nevertheless, in order to provide some
feeling as to the their size. As we will see shortly, the differences 
of entries in the third
column of Table III will provide the theoretical predictions 
in the fourth column in Table IV below.

Taking second order differences, we obtain
\beq{secdifflog}    
\begin{array}{ccc}
[E(ix) - E(jx)] - [E(iy) - E(jy)]  
&=& 
\\
\langle m_i-m_j \rangle_{xMes} - \langle m_i-m_j \rangle_{yMes} &=&
\,\,\ds {V_0\over2} \cdot 
\log \left [{{(m_i + m_x)\cdot (m_j + m_y)}\over{ (m_j + m_x)\cdot (m_i + m_y)}}
 \right ] 
\end{array}
\label{double_dif}
 \end{equation} 
We are now ready to see
to what extent the inclusion of the confining potential can reproduce the
observed reduction of effective quark mass differences with increase
of the spectator mass. To that effect, we
we examine the data and the theoretical values for 
the differences (\ref{secdifflog}). The results are given in Table IV
below.

\vskip1cm
\vbox{
\centerline{\bf TABLE IV}
\centerline{Comparison of experiment and theory for quark
double mass differences}
\def\baselinestretch{1.1}\small\normalsize
$$
\begin{array}{|c||c|c||c|c||c||c|}
\hline
&\multicolumn{5}{c||}{\rm Experiment}& \hbox{Theoretical}
\\
\cline{2-6}
&&&&&\,\,\,\, \langle m_i-m_j \rangle_{x}&
\hbox{prediction}\\
\raisebox{1.5ex}[0pt]{\hbox{observable}}& 
\,\,x \,\,& \raisebox{1.0ex}[0pt]{\hbox{$\langle m_i-m_j \rangle_{x}$}}&
\,\,y \,\,& \raisebox{1.0ex}[0pt]{\hbox{$\langle m_i-m_j \rangle_{y}$}}&
- \langle m_i-m_j \rangle_{y}  &
\hbox{eq.~(\ref{double_dif})}\\
&&{\rm MeV}&&{\rm MeV}&{\rm MeV}
&\hbox{MeV}\\
\hline
\hline
\medstrut \langle m_s-m_u \rangle & d & 179 & c & 103 & 76 & 51
\\ \hline
\medstrut \langle m_c-m_u \rangle & d & 1360 & c & 1095 & 265 & 202
\\ \hline
\medstrut \langle m_c-m_s \rangle & d & 1180 & c & 991 & 189 & 151 
\\ \hline
\medstrut \langle m_s-m_u \rangle & c & 103 & b & 91 & 12 &  18
\\ \hline
\end{array}
$$
}
The results in Table IV confirm the suggestion that the effective
mass of a bound quark depends upon its environment in a manner that is
approximated by a potential model. The results using the Quigg-Rosner potential 
are qualitatively correct. 
The first three rows of Table IV
show that  the data are all in
the direction of giving a larger difference than the predictions for cases
where mass differences  with a charm spectator are
compared with mass differences with a light spectator.
This is not true in the last row, where $c$-quark spectator is replaced by 
$b$-quark spectator,
and where the differences  12 MeV and 18 MeV are small.

In the next Section we will see how the quantitative agreement with the
data can be improved through the use of a more realistic potential, but
for the time being we 
apply the simple logarithmic potential to the case of $B_c$.

\subsection*{Application to the $B_{\MakeLowercase c}$ }
\noindent
The same approach applied to the $B_c$ mesons gives
\begin{equation}
\langle m_c-m_s \rangle_{bMes}= {{3(M_{B_c^{\scriptstyle *}}
-M_{B_s^{\scriptstyle *}})
+M_{B_c}-M_{B_s}}\over 4} =  {{3(M_{B_c^{\scriptstyle *}})
+M_{B_c}}\over 4} -
 {{3(
M_{B_s^{\scriptstyle *}})
+M_{B_s}}\over 4}
\end{equation}
The mass of the $B_c$ is known with large error as $ 6400 \pm 400 \,{\rm MeV}$
and the mass of the $B_c^{\scriptstyle *}$ is not known at all. We therefore
rearrange this relation to obtain a prediction,
\begin{equation}
 {{3(M_{B_c^{\scriptstyle *}})
+M_{B_c}}\over 4} = 
 {{3(
M_{B_s^{\scriptstyle *}})
+M_{B_s}}\over 4} + \langle m_c-m_s \rangle_{bMes}=
5314  \,{\rm MeV}+ \langle m_c-m_s \rangle_{bMes}
\end{equation}
To obtain $\langle m_c-m_s \rangle_{bMes}$, we note that
\begin{equation}
\langle m_c-m_s \rangle_{dMes}={{3(M_{D^{\scriptstyle *}}-
M_{K^{\scriptstyle *}})+M_D-M_K}\over 4} = 1178 \,{\rm MeV}.
\end{equation}
\begin{equation}
\langle m_c-m_s \rangle_{cMes}={{3(M_{\psi}-M_{D_s^{\scriptstyle *}})
+M_{\eta_c} - M_{D_s}}\over 4} = 992 \,{\rm MeV}.
\end{equation}

We correct these values using our results from the logarithmic potential to 
obtain,

\begin{equation}
\langle m_c-m_s \rangle_{bMes}=\langle m_c-m_s \rangle_{dMes} 
+ E(sd) - E(cd)  - E(bs) + E(bc) = 944 \,{\rm MeV}.
\end{equation}

This gives
\begin{equation}
M_{B_c} \leq {{3(M_{B_c^{\scriptstyle *}})
+M_{B_c}}\over 4} =
5314 + 944 \,{\rm MeV = 6258 } \,{\rm MeV}.
\end{equation}
where the inequality follows from the known sign of the unmeasured
hyperfine splitting. Alternatively we can use
\begin{equation}
\langle m_c-m_s \rangle_{bMes}=\langle m_c-m_s \rangle_{cMes}+
  E(sc) - E(cc)  - E(bs) + E(bc) = 909 \,{\rm MeV}.
\end{equation}
This gives
\begin{equation}
M_{B_c} \leq  {{3(M_{B_c^{\scriptstyle *}})
+M_{B_c}}\over 4} =
5314 + 909 \,{\rm MeV} = 6223 \,{\rm MeV}.
\end{equation}
Both predicted inequalities agree with experiment within the large
experimental errors.

\section {Use of different  potentials}  

In Section III we have shown that the logarithmic potential correctly 
reproduces the important qualitative empirical observation that 
the effective quark mass difference decreases when a lighter spectator quark
is replaced by a heavier spectator quark. 

Beyond this qualitative success, 
we note however that 
in the first three rows of Table IV, corresponding to 
$d\rightarrow c$, theory 
systematically undershoots the data, while in the fourth row,
corresponding to $c \rightarrow b$, theory overshoots the data slightly,
but in the latter case both experimental and theoretical differences are small.

That this discrepancy will be improved by using a more realistic potential can
be seen by noting from (\ref{FH1}) that the theoretical prediction is
proportional to an integral over the expectation value $\langle T \rangle$
of the kinetic energy. Since the commonly accepted combination of Coulomb plus 
linear potential increases much more sharply than the logarithmic potential at
small and large distances, the wave functions are confined to a smaller region
in space and the kinetic energy is therefore higher.   

To see this quantitatively we  consider the family of potentials which are
singular at the origin and confining at large distances.

\begin{equation}
V = {{V_o}\over{2n}}\left[\left({{r}\over{r_o}}\right)^n - 
\left({{r}\over{r_o}}\right)^{-n} \right]
\end{equation}
The value $n=1$ gives the combination of Coulomb plus linear  potential; the
value n=0 gives the logarithmic potential.  
The virial theorem then gives
\begin{equation}
\langle T \rangle = \left\langle {{r}\over{2}} \cdot {{dV}\over{dr}} 
\right\rangle = 
{{V_o}\over{4}}\left[\left({{r}\over{r_o}}\right)^n + 
\left({{r}\over{r_o}}\right)^{-n} \right]
\end{equation}
For $n > 0$ the kinetic energy $\langle T \rangle$ is seen to be no longer a
constant but  dependent upon powers of the radial co-ordinate $r$. For a
logarithmic potential $r$ scales\cite{quiggros} with the reduced mass $m_r$
like $m_r^{-1/2}$. Thus $\langle T \rangle$ will be larger in lighter quark
systems than in the  $c \bar c$ system for which the value of the parameter 
$V_o$ was defined to  fit charmonium data. The values of the integral
(\ref{FH1}) will therefore be  larger than the predictions from the logarithmic
potential  model.

\pagebreak
 \section{The dynamics of a diquark-triquark pentaquark}
  
The recent observation observation of the strange $\Theta^+$
 pentaquark\cite{Kyoto,Russia,Stepanyan:2003qr} with a mass of 1540 MeV
and a very small width $\sim$20 MeV
has generated a great deal of 
interest. Although the original prediction of an exotic $KN$ resonance was 
obtained within the framework of the Skyrme model \cite{SMKNa,SMKNb}, there is
an obvious and urgent need to understand what $\Theta^+$ is in the quark language
\cite{Stancu:2003if}.

The first guess is to use the Sakharov-Zeldovich approach, together with
Jaffe's color-magnetic interaction\cite{Jaffe} model for the hyperfine
interaction. Previous attempts show, however, that a single-cluster description
of the ($uudd\bar s$) system fails because the  color-magnetic repulsion
between flavor-symmetric states  prevents binding.

An additional nontrivial challenge for the quark interpretation \cite{JaffePC} 
is that whereas the Skyrme model predicts
that $\Theta^+$ has positive parity, the ``standard" pentaquark involves 5 quarks
in an $S$-wave and therefore has negative parity. As of now, there is no clearcut
experimental information on the $\Theta^+$ parity, but if it is positive,
clearly one must have one unit
of orbital angular momentum and this makes the calculation difficult.

We consider here a possible model for a strange pentaquark that bypasses these
difficulties by dividing the system into two color non-singlet clusters 
which separate the pairs of identical flavor.
The two clusters, a $ud$ diquark and a $ud\bar s$ triquark,
are separated by a distance larger than the range of the color-magnetic force 
and are kept together by the color electric force. Therefore the color hyperfine
interaction operates only within each cluster, but is not felt between the
clusters.

The $ud$ diquark is in the $\bar {\hbox{\bf 3}}$
of the color $SU(3)$  and in  the $\bar {\hbox{\bf 3}}$ of the flavor $SU(3)$  
and has $I=0, S=0$, like the $ud$
diquark in the $\Lambda$. It is in the symmetric {\bf 21} of the color-spin $SU(6)$ 
and is antisymmetric in both spin and color. 
 
The {\bf 21} representation of $SU(6)$ contains a color antitriplet with spin 0
and a color sextet with spin 1.  

The $ud$ in the $ud\bar s$ triquark is in {\bf 6} of $SU(3)_c$,
in $\bar{\hbox{\bf 3}}$ of $SU(3)_f$ and has $I=0, S=1 $.  
It is also in the symmetric 
{\bf 21} of the color-spin $SU(6)$, but is symmetric in both spin and color.   

The triquark consists of the diquark and antiquark coupled to an $SU(3)_c$ 
triplet and has $I=0, S=1/2$. It is in the fundamental {\bf 6} representation of
the color-spin $SU(6)$. It is in a $\bar{\hbox{\bf 6}}$ of  $SU(3)_f$.   

We now define the classification of the diquarks with spin $S$,
denoted by $\ket{(2q)^S}$ and the triquark, denoted by  
 $\ket{(2q\bar s)^{1\over2}}$,
in a conventional notation $\ket{D_6,D_3,S,N}$
\cite{Patera,Sorba}
where $D_6$ and $D_3$ denote the dimensions of the color-spin $SU(6)$ and
color $SU(3)$ representations in which the multiquark states are classified,
$S$ and $N$ denote the total spin and the number of quarks in the system,
\def\smallhalf{\hbox{${1\over2}$}}
\begin{eqnarray}
\ket{\,\,(2q)^1\,}          &=& \ket{21,6,1,2}    \cr
\ket{\,\,(2q)^0\,}          &=& \ket{21,\bar 3,0,2}   \\
\ket{(2q\bar s)^{1\over2}}    &=& \ket{\,\,6,3,\smallhalf,3}    \nonumber
\end{eqnarray}

A standard treatment using the $SU(6)$ color-spin algebra\cite{Patera,Sorba}
gives the result that  the hyperfine interaction is stronger by
${1\over6}(M_\Delta-M_N)$ for the diquark-triquark system than for the kaon
nucleon system,
\begin{equation}
[V(2q\bar s^{1\over2}) + V(2q^0)] - [V(K )+ V(N)]=
-{1\over6}(M_\Delta-M_N)  \approx 50 {\rm MeV}
\label{MassShift}
\end{equation}
The physics here is simple. The spin-zero diquark is the same as the diquark in
a $\Lambda$ and has the same hyperfine energy as a nucleon. A triquark with
one  quark coupled with the $\bar s$ antiquark to spin zero has the same
hyperfine energy as a kaon but no interaction with the other quark. The triquark
coupling used here allows the $\bar s$ antiquark to interact with both the $u$
and $d$ quarks and gain hyperfine energy with respect to the case of the kaon.
For an isolated triquark such a configuration is of course forbidden, since it a
color nonsinglet, but here it is OK, since the the triquark color charge
is neutralized by the diquark.

We see that the triquark-diquark system will be somewhat more bound than a kaon
and a nucleon. The diquark and triquark will have a color electric interaction
between them which is identical to the quark-antiquark interaction in a meson.
If we neglect the finite sizes of the diquark and triquark we can compare
this system with analogous mesons. We can use the effective quark masses that
fit the low-lying mass spectrum 
\beq{qmass}
m_u =  360  
\hbox{\ MeV};\qquad
m_s=  540 
\hbox{\ MeV};\qquad
m_c= 1710 
\hbox{\ MeV}\,.
\end{equation}
we find a very rough estimate 
\beq{qmass2}
m_{diq} =  720  
\hbox{\ MeV};\quad
m_{triq}=  1260
\hbox{\ MeV};\quad
m_r(di\hbox{-}tri)= 458
\hbox{\ MeV};\quad
m_r(c\bar s)=  410
\hbox{\ MeV}\,.
\end{equation}
where $m_{diq}$ and
$m_{triq}$ denote the  
effective masses of the diquark and triquark, $m_r(di\hbox{-}tri)$ denotes the reduced 
mass for the relative motion of the diquark-triquark system and  
$m_r(c\bar s)$ denotes the reduced mass of the $c\bar s$ system used to describe the 
internal structure of the $D_s$ spectrum.

A crucial observation is that the diquark-triquark system may not exist in 
a relative $S$-wave. This is because in $S$-wave the hyperfine interaction acts not
only within the clusters but also between them. The repulsive terms may then win and
the would be $S$-wave gets rearranged into the usual $K N$ system. The situation
is different in a $P$-wave, because then the diquark and the triquark are separated
by an angular momentum barrier and the color-magnetic interactions operate only
within the two clusters.  The price is the $P$-wave excitation energy.

We can obtain a rough estimate of this $P$-wave excitation energy, using the fact
that the reduced mass of the $D_s$ is close to the reduced mass of the 
diquark-triquark system and that all the relevant experimental information about the 
$D_s$ system has recently become available.

It has been proposed that the new $D_s$(2317) \cite{Aubert:2003fg,CLEO,BELLE-Ds}
is a $0^+$ excitation \cite{Bardeen:2003kt}
of the ground state $0^-$ $D_s$(1969). If so, the \,350 MeV excitation energy then
consists of a $P$-wave contribution, on top of a contribution from color hyperfine
splitting. We can estimate the net $P$-wave
excitation energy \,$\delta E^{P-wave}$\,
by subtracting the $c$-$s$ hyperfine splitting
obtained from the mass difference between $D_s^*$ and $D_s$,
\begin{equation}
\delta E^{P-wave} \approx  350 -
(m_{D_s^*} - m_{D_s}) = 207\ \rm MeV
\end{equation}

From eq.~(\ref{MassShift}) we infer that without the $P$-wave excitation
energy the diquark-triquark mass is
$m_N+m_K-{1\over6}(M_\Delta-M_N)\approx 1385$ MeV, 
so that the total mass of the 
$P$-wave excitation of the diquark-triquark system is
is expected to be 
\beq{PentaMass}
M_{di\hbox{-}tri} \approx 1385+207 = 1592 \ \rm MeV\,,
\eeq
about 3\% deviation from the observed mass of the $\Theta^+$ particle.
It should be kept in mind, however, that this is only a very rough
qualitative estimate and this close agreement might well be fortuitous, as
there are several additional model-dependent effects which should be 
taken into account:
the reduced mass of $D_s$ is $\sim 12\%$ lower than $m_r(di\hbox{-}tri)$,
we don't know the spatial wave functions and we
have neglected the spatial extent of the
diquark and triquark and possible molecular Van-der-Waals interactions
spatially polarizing the two, breaking of flavor $SU(3)$, etc.

In addition to the parity and the mass, we also note that our model
naturally gives a state with isospin zero
because both the diquark and triquark have $I=0$. The isospin has not yet been
determined experimentally, but no isospin partners of the $\Theta^+$ have been
found and the Skyrme also predicted $I=0$. 
This should be contrasted 
with attempts to envision the $\Theta^+$ as a $KN$ molecule 
in a $P$-wave \cite{Mitya}, which  have a problem in getting rid of the $I=1$ state. 

Our model also naturally predicts that the $\Theta^+$ is in
an antidecuplet of $SU(3)$ flavor. The diquark is a $\bar{\hbox{\bf 3}}$, the 
triquark a $\bar{\hbox{\bf 6}}$ and 
in $SU(3)$ 
$\bar{\hbox{\bf 3}} \otimes \bar{\hbox{\bf 6}}=
\overline{\hbox{\bf 10}} \oplus  {\hbox{\bf 8}}$
and only $\overline{\hbox{\bf 10}}$ has the right strangeness. 
$KN$ is $\hbox{\bf 8} \otimes \hbox{\bf 8}$
in $SU(3)_f$ and contains {\bf 27} with an isovector with the
right strangeness, in addition to an antidecuplet.
The antidecuplet prediction is again 
in agreement with the Skyrme model.

Since $M_{di\hbox{-}tri}$ is above the $KN$ threshold, 
the system will eventually decay to 
$KN$, but the orbital angular momentum barrier and
the required color rearrangement will make such a decay relatively slow, 
possibly explaining the observed narrow width of the $\Theta^+$.

\section*{Summary and Conclusions}
In summary, we have demonstrated a large class of simple phenomenological
hadronic mass relations. We use them to construct a prediction 
involving the only roughly known mass of $B_c$ and the 
yet unmeasured mass of $B_c^*$.
Direct derivation of such relations from QCD is still
an open challenge. 

The simple generalized
Sakharov-Zeldovich mass formula holds with a single set of effective quark mass
values for all ground state mesons and baryons having no more than one  strange
or heavy quark. The breakdown of this simple description in mesons having two
strange or heavy quarks implies that the effective mass of heavy quarks
decreases sharply when the companion light quark is replaced by a heavy
one.  We are able to reproduce this phenomenon 
qualitatively through the increased binding of
heavier quarks in the short-range Coulomb-like part of the potential and
semiquantitatively in a crude potential model. 

Using these ideas, we propose the interpretation 
of the recently discovered $\Theta^+$ exotic $KN$ resonance
as a novel kind of a pentaquark, involving a recoupling
of the five quarks into a diquark-triquark system in non-standard color 
representations. 
We provide a rough numerical estimate indicating that such a color recoupling 
might put the pentaquark mass in the right ballpark of the experimentally observed
$\Theta^+$ mass. Our model naturally predicts that $\Theta^+$ has spin 1/2,
positive parity, is an isosinglet and is an antidecuplet in $SU(3)_f$.

Regardless of the specific details of the model, we have addressed the
problem what kind of a five-quark configuration can describe
the $\Theta^+$. We have shown that our new diquark-triquark model with color 
recoupling gives a lower mass than than the simplest
$uudd\bar s$ and it looks promising. The diquark-triquark configuration
might also turn out to be useful if negative parity 
exotic baryons are experimentally discovered in future.

\pagebreak
\section*{Acknowledgments}

The research of one of us (M.K.) was supported in part by a grant from the
United States-Israel Binational Science Foundation (BSF), Jerusalem and
by the Einstein Center for Theoretical Physics at the Weizmann Institute.
We benefited from e-mail discussions with Ken Hicks about the experimental data on
the $\Theta^+$ and
with Simon Capstick, Frank Close, Mitya Diakonov, Bob Jaffe and Micha{\l}
Prasza{\l}owicz about the challenges this data poses for a theoretical
interpretation.

%
\catcode`\@=11 
\def\references{ 
\ifpreprintsty \vskip 10ex
%
\hbox to\hsize{\hss \large \refname \hss }\else 
\vskip 24pt \hrule width\hsize \relax \vskip 1.6cm \fi \list 
{\@biblabel {\arabic {enumiv}}}
{\labelwidth \WidestRefLabelThusFar \labelsep 4pt \leftmargin \labelwidth 
\advance \leftmargin \labelsep \ifdim \baselinestretch pt>1 pt 
\parsep 4pt\relax \else \parsep 0pt\relax \fi \itemsep \parsep \usecounter 
{enumiv}\let \p@enumiv \@empty \def \theenumiv {\arabic {enumiv}}}
\let \newblock \relax \sloppy
 \clubpenalty 4000\widowpenalty 4000 \sfcode `\.=1000\relax \ifpreprintsty 
\else \small \fi}
\catcode`\@=12 
{\tighten

}

\end{document}